\documentclass[aps,twocolumn,a4paper,showpacs,superscriptaddress]{revtex4}

\usepackage{graphicx}
\usepackage{amsmath}
\usepackage{amssymb}
\usepackage{enumerate}
\usepackage{subfigure}
\usepackage{tabularx}

\newcommand{\be}{\begin{equation}}
\newcommand{\ee}{\end{equation}}
\newcommand{\ben}{\begin{eqnarray}}
\newcommand{\een}{\end{eqnarray}}
\newcommand{\bes}{\begin{subequations}}
\newcommand{\ees}{\end{subequations}}

\newcommand{\sech}{{\rm sech}}
\newcommand{\LL}{{\cal L}}

\begin{document}

\title{Highly interactive kink solutions}
\author{A. R. Gomes}
\affiliation{Institut f\"ur Theoretische Physik,
Universit\"at Heidelberg, Philosophenweg 16,
D-69120 Heidelberg, Germany.}
\affiliation{Departamento de F\'\i sica, Instituto Federal do Maranh\~ao, 65030-000 S\~ao Lu\'\i s, MA, Brazil}
\author{R. Menezes}
\affiliation{Departamento de Ci\^encias Exatas, Universidade Federal
da Para\'{\i}ba, 58297-000 Rio Tinto, PB, Brazil}
\affiliation{Departamento de F\'\i sica, Universidade Federal de Campina Grande, 58109-970, Campina Grande, Para\'\i ba, Brazil}
\author{J. C. R. E. Oliveira}
\affiliation{Centro de F\'\i sica do Porto, Rua do Campo Alegre 687, 4169-007 Porto, Portugal}
\affiliation{Departamento de Engenharia F\'\i sica da Faculdade de Engenharia da Universidade do Porto, Rua Dr. Roberto Frias, s/n, 4200-465 Porto, Portugal}

\begin{abstract}

{In this work we present a new class of real scalar field models admitting strongly interactive kink solutions.  Instead of the usual exponential asymptotic  behavior these topological solutions exhibit a power-law one. We investigate the interaction force between a pair of kink/anti-kink solutions both analytically and numerically, by integrating the time dependent field equations of the model. Furthermore, working within the first-order framework, we  analyze the linear stability of these solutions. The stability analysis leads to Sch\"odinger-like  equations with potentials which, despite admitting no bound states, lead to strong resonance peaks. We argue that these properties are important for some possible physical applications}.

\end{abstract}
\date{\today}
\pacs{11.27.+d, 11.10.Lm}
\maketitle
\section{Introduction\label{Intro}}

Topological structures associated with solitary wave solutions of non-linear field theories are of great interest in several areas of physics \cite{Vachaspati:2006zz,VSh,ManSut}. Particular topological {$(3,1)$-dimensional} kink-like defects, associated with interfaces between distinct order parameter regions, have been studied in detail both in non-relativistic and relativistic contexts. Phase-field models have been {successfully} applied { for describing} non-relativistic interface dynamics across many fields in condensed matter  and material science systems \cite{GW,Provatas,WH,Travasso,loginova,bray}. The dynamics of relativistic interfaces has also extensively been investigated, mainly in the context of cosmology, where domain wall networks that should have been formed in primordial phase transitions are considered, along with their cosmological implications \cite{wall1,wall2,wall3,wall4}. Recently a unified theoretical paradigm for interface dynamics, which includes both relativistic and non-relativistic systems in a unified framework, has been proposed in \cite{Avelino:2010qf}. Moreover,  in {$(4,1)$-dimensional} braneworld theories, one can construct thick Minkowski branes which splits in a {warped bulk} spacetime due to a fist-order transition in a {warped bulk} \cite{acampos}.

In the present work we focus our attention on kink-type solutions, as topological static solutions of models described by a single real scalar field in bidimensional spacetime. Relativistic scalar fields models, admitting a large variety of kink-type solutions, have been studied in a diversity of scenarios \cite{DeWolfe,Toharia:2007xe,Basar:2008im,Melfo:2011ev,VBA,AlonsoIzquierdo:2010ze,ABC,Bazeia:2006pj}.  Models allowing different kink/anti-kink interaction  behavior \cite{AKL,CampSch,AOM,doi, Dorey:2011yw} have been motivated both theoretically and experimentally.  Peculiar kink dynamics have also been studied in the context {of models with modified kinetic term}  \cite{Babichev:2006cy, Bazeia:2007df,Adam:2007ij,Bazeia:2008tj,Bazeia:2011sb,Adam:2011jt}

We introduce a new model with the particular feature of having a null second-order derivative of the potential at the minima. As it is known, the second-order derivative of the potential is related with the mass scale of the quantum meson, associated with the frequency of oscillation of the field at the minimum of the potential. A linear perturbation $\eta$ of a vacuum solution, $\phi=v+\eta(x)$, obeys the fluctuation equation  $\Box \eta + m_v^2\eta=0$.  The perturbation solution, $\eta=\cos(\omega t - kx)$,  oscillates with an energy $\omega^2=k^2+m_v^2$ where the meson mass is given by $m_v^2=V^{(2)}_{\phi}(v)$ (the notation $V_\phi^{(n)}=d^n V(\phi)/d\phi^n$, { for integer $n$ is being used}). In models admitting kink solutions, { with $V_{\phi}^{(2)}(v)=m_v^2\neq0$, the asymptotic solution} is given by $\phi(x)\approx v+\phi_*(x)$ where
 \be
\frac{d^2\phi_*}{dx^2}=m_v^2 \phi_*
 \ee
and $\phi_*(x)$ has an exponential  behavior $\phi_*(x)\propto e^{-m_v \,x}$ when $x\to \infty$. When $V^{(n)}_{\phi}(v)=0$, all the linear perturbation modes are massless and the solutions asymptotic  behavior is related to a higher order of the potential derivative at the minima. It is possible to show that for a model, in which the first $n$ order derivatives of the potential vanish at the minimum, the equation of motion for the asymptotic kink solution is given by
\be
\frac{d^2\phi_*}{dx^2} =  a_{n+1} \phi^{n}_*,
\ee
where $a_n=V^{(n)}_\phi(v)$, allowing a power-law  behavior of  model's asymptotic solutions,
\be\label{powerlaw}
\phi_* =\frac{A_n} {x^{\frac{2}{n-1}}},
\ee
where the constant $A_n$ depends on $n$ and $a_n$.

In this paper we introduce the smooth minima potential model corresponding to $n=3$. The  power-law behavior of this model asymptotic solution, and corresponding energy density, originates a non-standard kink/anti-kink interaction pattern, that is analyzed in detail in this work. We also investigate the generalized { $n$-th} dependent class of models {finding for the fluctuations  narrow resonance peaks whose lifetime gets reduced with the increasing of $n$}.

The paper is organized in the following way: In section \ref{framework} we explore the framework of the scalar fields theories that support kink solutions. In section \ref{themodel} we introduce the model and discuss its  specific properties. In section  \ref{stability} we investigate the linear stability of the model solutions.  In section \ref{generalization} { we present the generalized class  for these models  and we discuss resonance effects}. We end the work in section \ref{end} presenting some comments,  conclusions {and perspectives}.

\section{The framework\label{framework}}

Let us consider the standard Lagrangian density
\be\label{full_eq}
\LL = \frac12 \partial_\mu \phi \partial^\mu \phi - V(\phi),
\ee
where $\phi$ is a real scalar field and $V(\phi)$ is the potential which specifies the model under consideration, $x^\mu=(t,x)$ and $x_\mu=(t,-x)$. We consider that the potential engenders a set of critical points, $v_i=\{\bar\phi_1,\bar\phi_2,\ldots,\bar\phi_n\}$, such that $V^{(1)}_\phi(v_i)=0$ and $V(v_i)=0$ for $i=1,2,\ldots,n$.

The equation of motion following from the above model is
\be\label{eqmotion222}
\frac{\partial^2 \phi}{\partial t^2} -
\frac{\partial^2 \phi}{\partial x^2} +V^{(1)}_\phi = 0.
\ee

The corresponding components of the  energy-momentum tensor are given by
\bes
\ben
\rho(x)=T_{00}&=&\frac{1}{2}\left(\frac{d\phi}{dt}\right)^2\! + \frac{1}{2}\left(\frac{d\phi}{dx}\right)^2\! + V(\phi), \\
T_{01}&=&T_{10} =\left(\frac{d\phi}{dt}\right)\left(\frac{d\phi}{dx}\right),\\
p=T_{11}&=&\frac{1}{2}\left(\frac{d\phi}{dt}\right)^2\! + \frac{1}{2}\left(\frac{d\phi}{dx}\right)^2\! - V(\phi).
\een
\ees

For static solutions $\phi=\phi(x)$, we can write the equation of motion as
\be
\frac{d^2\phi}{dx^2}=V_\phi^{(1)}.
\ee
Integrating the equation of motion we obtain
\be\label{equationfirst}
\frac{1}{2}\left(\frac{d\phi}{dx}\right)^2 = V(\phi)+C
\ee
where $C$ is an integration constant, identified as the solution pressure. The topological solutions necessarily obey the boundary conditions
\be
\lim_{x\to\pm\infty} \frac{d\phi}{dx} \to 0
\ee
in order to ensure finiteness of the gradient portion of the energy. These solutions are stressless, guaranteeing $C=0$ and $V(v_i)=0$ for the $v_i$ set of potential minima. For $C=0$ and $V(\phi)\geq0$, Eq. \eqref{equationfirst}  is then given by
\be\label{firstorder}
\frac{d\phi}{dx}= \sqrt{2V(\phi)}\,\,\,\,\,{\rm and} \,\,\,\,\, \frac{d\phi}{dx}=- \sqrt{2V(\phi)}.
\ee
The first (second) equation supports the monotonically  increasing  (decreasing) solution connecting two adjacent minima, called kink (anti-kink).  For simplicity we will restrict ourselves to the explicit kink-type solution. The anti-kink solution  may be obtained by the transformation $\bar\phi(x)=\phi(-x)$.

The superpotential function $W(\phi)$ may be defined as
\be
W(\phi)=\int^\phi \sqrt{2V(\tilde\phi)}\, d\tilde\phi,
\ee
allowing to express the energy of the solution as
\be
E=\int_{-\infty}^\infty \rho(x)=|W(\phi(\infty))-W(\phi(-\infty))|
\ee
without knowing the explicit form of the solution.

The first order equations \eqref{firstorder}, for the kink solution, may then be written as \cite{Bog,PS}
\be
\frac{d\phi}{dx}= W^{(1)}_{\phi}.
\ee

A model extensively studied in various scenarios \cite{Vilenkin:1994pv,DeWolfe,loginova,Dinda} is the well-known double-well potential given by
\be\label{phi4model}
V_s(\phi)=\frac{\tilde\lambda}2 (b^2-\phi^2)^2,
\ee
where $b$ and $\tilde \lambda$  are positive parameters. $V^{(2)}_\phi(\pm b)=4\tilde\lambda b^2$
The superpotential function for this model is then given by
\be\label{Wphi4}
W_s(\phi)=\sqrt{\tilde\lambda} \left(b^2\phi-\frac13 \phi^3\right),
\ee
leading to the static solution
\be\label{kinkp4}
\phi_s(x)=b\tanh(b\,\sqrt{\tilde\lambda}  \,x),
\ee
{and with} an energy density given by
\be
\rho_s(x)=\tilde\lambda b^4 \sech^4(b\,\sqrt{\tilde\lambda} \,x).
\ee
The corresponding total energy of the solution is
\be
E_s=\frac{4b^3\sqrt{\tilde\lambda}}{3}.
\ee
The thickness of solution  \eqref{kinkp4} is defined by the following condition $\delta_s =2(b\sqrt{\tilde\lambda})^{-1}$, such that
\be
\frac{\phi_s((b\sqrt{\tilde\lambda})^{-1})}{\phi_s(\infty)}=\tanh(1) \approx 0.762.
\ee
The solution is fully characterized by its energy and thickness.
\section{The model\label{themodel}}

\begin{figure}[h!]
  \includegraphics[width=6cm]{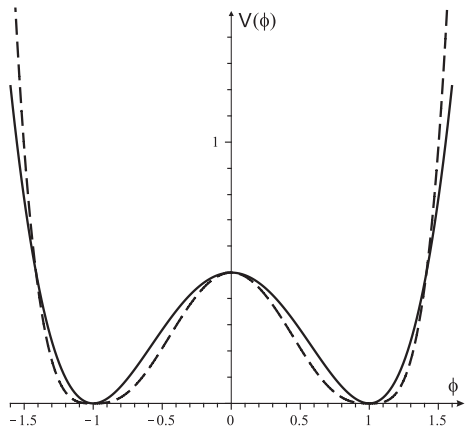}
  \caption{Potential profile of model \eqref{phi4model} (solid line) and model \eqref{newmodel} (dashed line) for $a=b=\tilde \lambda=\lambda=1$.   \label{pot1}}
     \label{fig:1}
  \end{figure}

We introduce the following model
\be\label{newmodel}
V(\phi)=\frac\lambda2 |a^2-\phi^2|^3,
\ee
where $a$ and $\lambda$  are positive parameters. {The model exhibits a  $Z_2$ symmetry $\phi \to -\phi$, with a local} maximum point in $\phi=0$, for $V(0)={\lambda a^6}/{2}$, and two minima points at $\phi=\pm a$. It is possible to observe in Fig. \ref{pot1}, that the two potential minima, with the particular characteristic of having a null potential second-order derivative $V^{(2)}_{\phi}(\pm a)=0$, are smoother than in \eqref{phi4model} model.

The model superpotential function is given by
\be\label{Wphi3}
W(\phi)\!=\!\frac{\sqrt{\lambda}a^4}8\!\begin{cases}\displaystyle\!
-f(\phi)- 3 {\rm arccosh}\left(\frac{|\phi|}{a}\right)-\frac{3\pi}2 & \phi<-a \\ \displaystyle
f(\phi)+ 3 \arcsin\left(\frac{\phi}{a}\right) & |\phi|\leq a \\ \displaystyle
-f(\phi)+ 3 {\rm arccosh}\left(\frac{|\phi|}{a}\right)+\frac{3\pi}2& \phi>a
\end{cases}
\ee
with
\be
f(\phi)= \frac{1}{a^4}\phi(5a^2-2\phi^2)\sqrt{|a^2-{\phi}^2|}.
\ee
The values of the superpotential function at the minima are given by $W(\pm a)=3{\sqrt{\lambda}a^4}/{16}$.

A similar model was presented in Ref \cite{Bazeia:2002xg}, as a deformation of $\phi^4$ model \eqref{phi4model}, however in that case the potential function is positive defined and the corresponding kink and anti-kink solutions connect two inflection points.

 The equation of motion of model \eqref{newmodel} is given by
\be\label{fullequation_new}
\partial_\mu \partial^\mu  \phi - 3\lambda\phi(a^2-\phi^2)|a^2-\phi^2|=0,
\ee
with the corresponding static solutions, $\phi=\phi(x)$, obeying the following equation
\be
\frac{d^2\phi}{dx^2}=- 3\lambda\phi(a^2-\phi^2)|a^2-\phi^2|.
\ee
The solution with finite energy obeys the BPS first order equations
\be
\frac{d\phi}{dx}= \sqrt{\lambda}(\sqrt{a^2-\phi^2})^3,
\ee
and the model topological solutions are then given by
\be\label{kink2}
\phi(x)=\frac{x \,a^3 \sqrt{\lambda}}{\sqrt{1+\lambda a^4 x^2}},
\ee
with a corresponding energy density
\be
\rho(x)=\frac{a^6 {\lambda}}{(1+\lambda a^4 x^2)^3}.
\ee
The total solution energy may then be calculated as
\be\label{E_0_new}
E=\frac{3}{8}a^4 \sqrt{\lambda} \pi.
\ee
Defining the solution thickness by $\delta  =2 (a^2\sqrt{\lambda})^{-1}$ we obtain

\be
\frac{\phi((a^2\sqrt{\lambda})^{-1})}{\phi(\infty)}=\frac{\sqrt{2}}{2} \approx 0.701,
\ee
validating this thickness definition.

\begin{figure}[h!]
  \includegraphics[width=6.2cm]{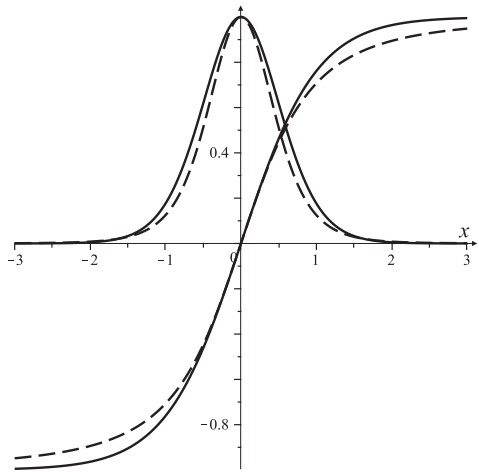}
   \caption{The $\phi^4$ kink solution profile and respective energy density, solid line, and the new-kink (\ref{kink2}) solution and energy density profile, dashed line, for $a=b=\tilde \lambda=\lambda=1$.   \label{kinks1}}
  \end{figure}

In Fig. \ref{kinks1}  the profile of the kink solutions (\ref{kinkp4}) and (\ref{kink2}) are plotted for comparison. It is possible to observe the slower asymptotic behavior of the new kink solution when compared with the  $\phi^4$ kink solution. In fact, for large positive values of $x$, solution \eqref{kink2} may be expanded, originating the polynomial behavior
\be
\phi(x)=a\left[1-\frac{1}{8} \left(\frac{\delta}{x}\right)^2+\frac{3}{128} \left(\frac{\delta}{x}\right)^4+\ldots\right],
\ee
while, as it is known, the $\phi^4$ model exhibits an asymptotic exponential behavior
\be
\phi_s(x)=b\left[1-4e^{-{x}/{\delta_s}}+2e^{-8{x}/{\delta_s}}+\ldots\right].
\ee

Note that the slower asymptotic behavior of the new kink solution is an intrinsic characteristic of the model, regardless of its solution thickness. The condition $V^{(2)}_{\phi}(\phi(x\to\pm\infty))=0$ is associated with the asymptotic polynomial behavior, which is always slower than the exponential one obtained for $V^{(2)}_{\phi}(\phi(x\to\pm\infty))\neq0$.

The new-kink energy density, shown in Fig. \ref{kinks1}, also exhibits a polynomial behavior
\be
\rho(x)=\frac{a^6\lambda}{64} \left(\frac{\delta}{x}\right)^6\left[1-\frac{3}{4} \left(\frac{\delta}{x}\right)^2 +\ldots\right], \,\, |x|\gg1,
\ee
while the energy density of the $\phi^4$ solution presents the usual exponential behavior,
\be
\rho_s(x)=16b^4\tilde\lambda e^{-8{x}/{\delta_s}}\left[1-4e^{-4{x}/{\delta_s}} +\ldots\right], \,\, |x|\gg1.
\ee

\begin{figure}[h!]
  \includegraphics[width=4.25cm]{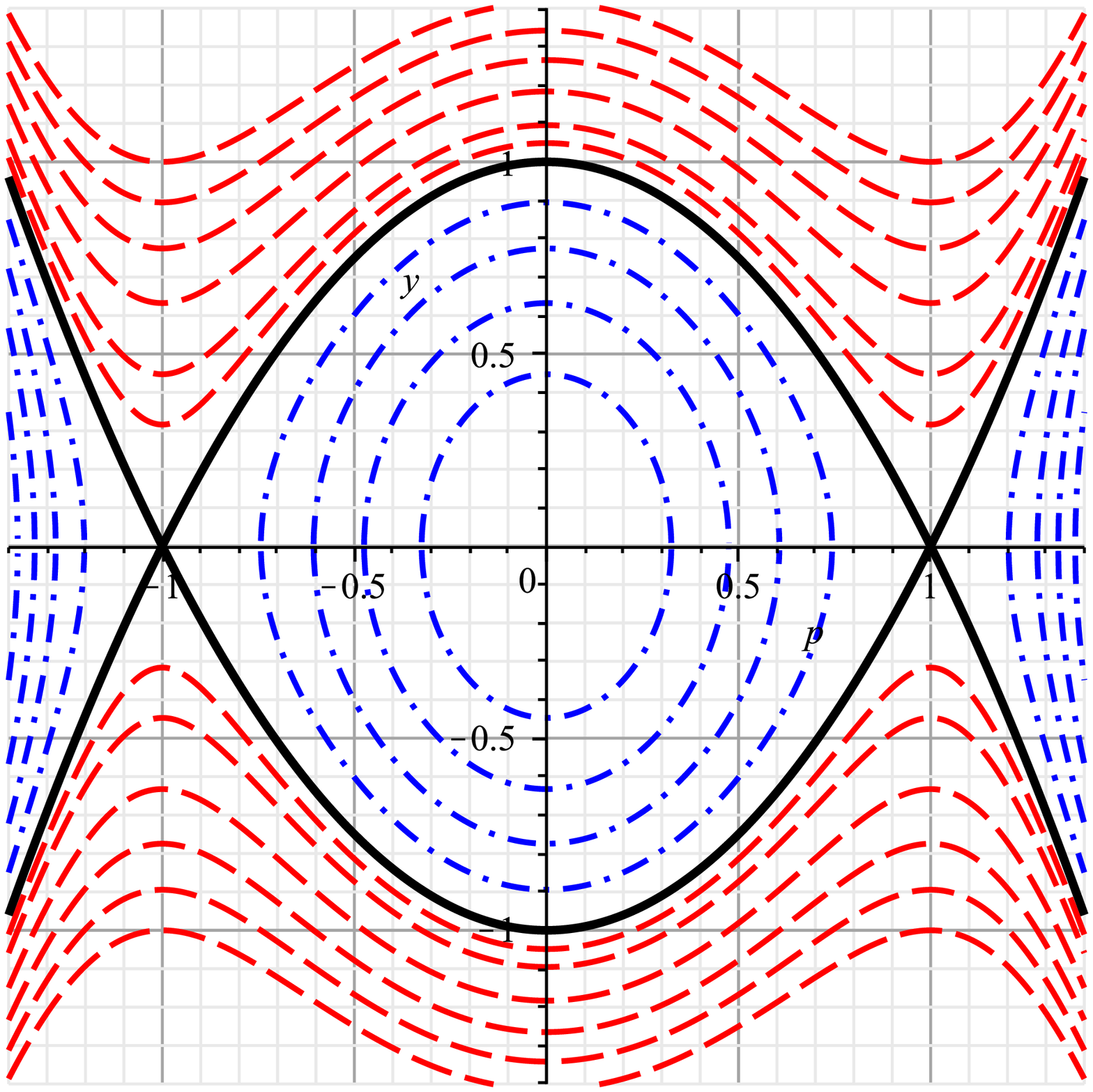}
  \includegraphics[width=4.25cm]{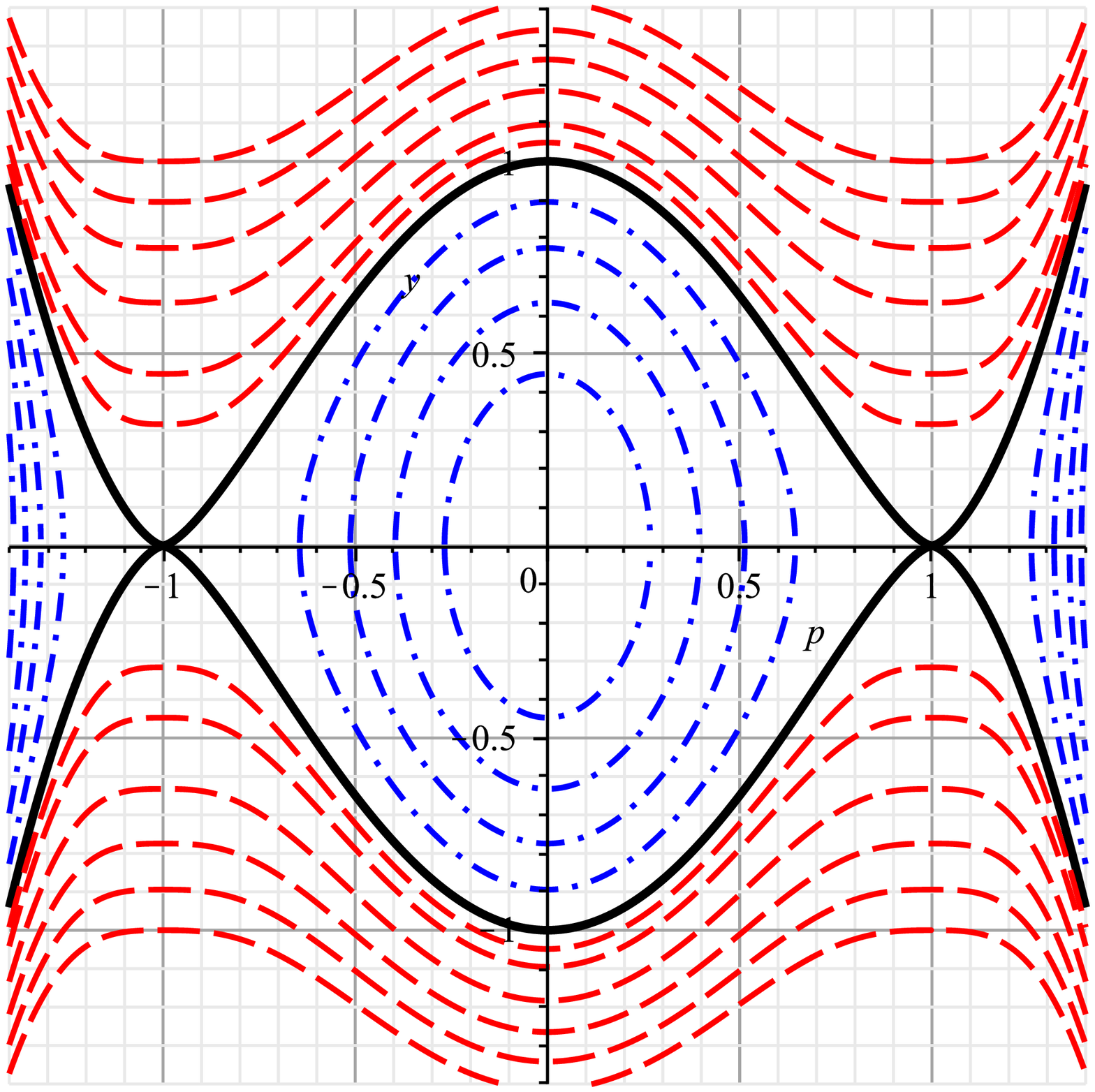} \caption{{Solution in the} $(\phi,d\phi/dx)$ plane for the $\phi^4$ model \eqref{phi4model} (left panel) and the new model \eqref{newmodel} (right panel). The solutions for positive pressure values are represented in blue (dash-dotted lines), for negative pressure values in red (dashed lines), and for a vanishing pressure the solution is represented in black (solid lines). \label{phase}}
  \end{figure}

For completeness, in Fig. \ref{phase}, the model solutions associated with non-vanishing pressure values are presented in the $(\phi,d\phi/dx)$ plane.

\begin{figure}[h!]
  \includegraphics[width=8.5cm]{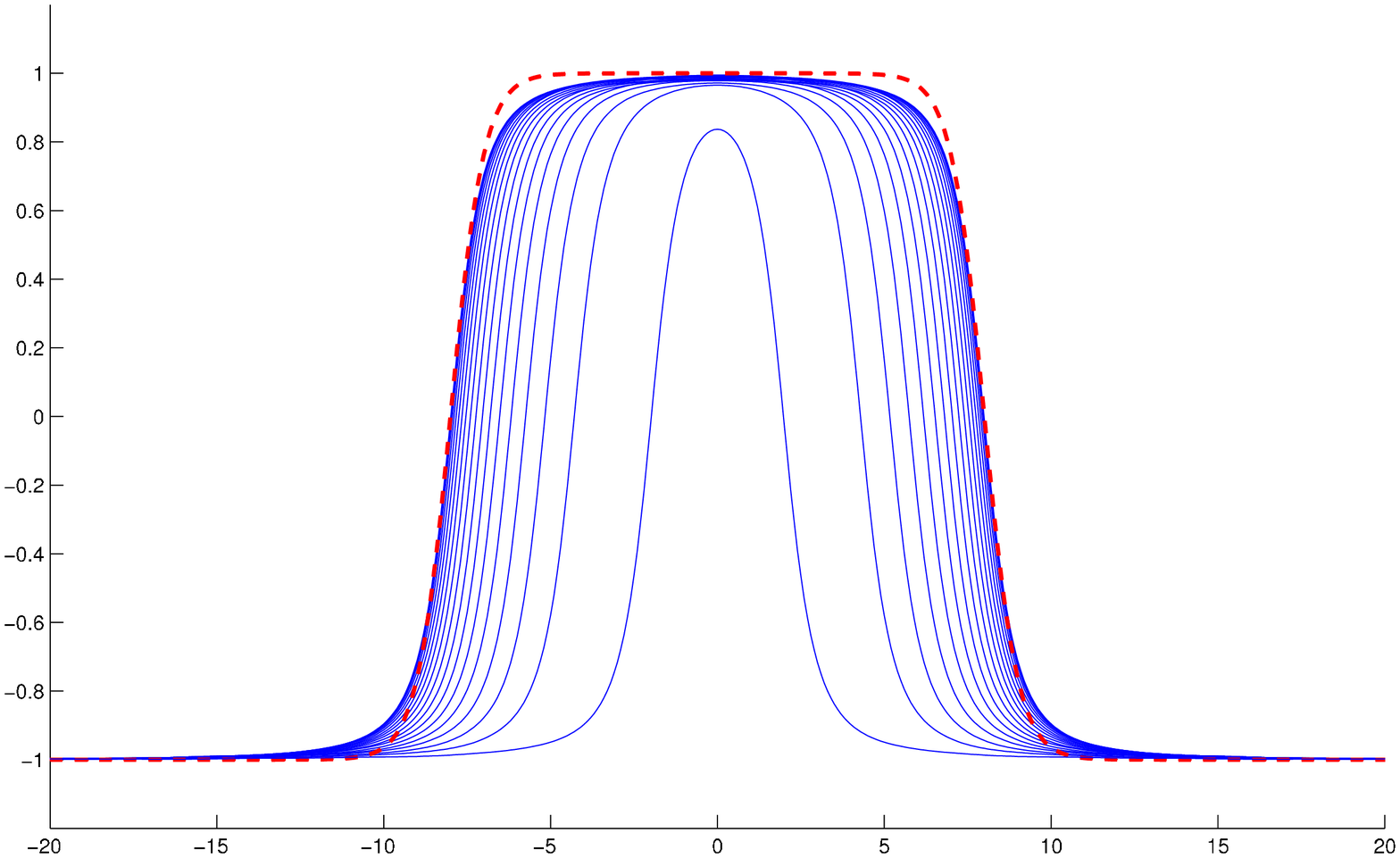}

  \vspace{0.4cm}

  \includegraphics[width=8.5cm]{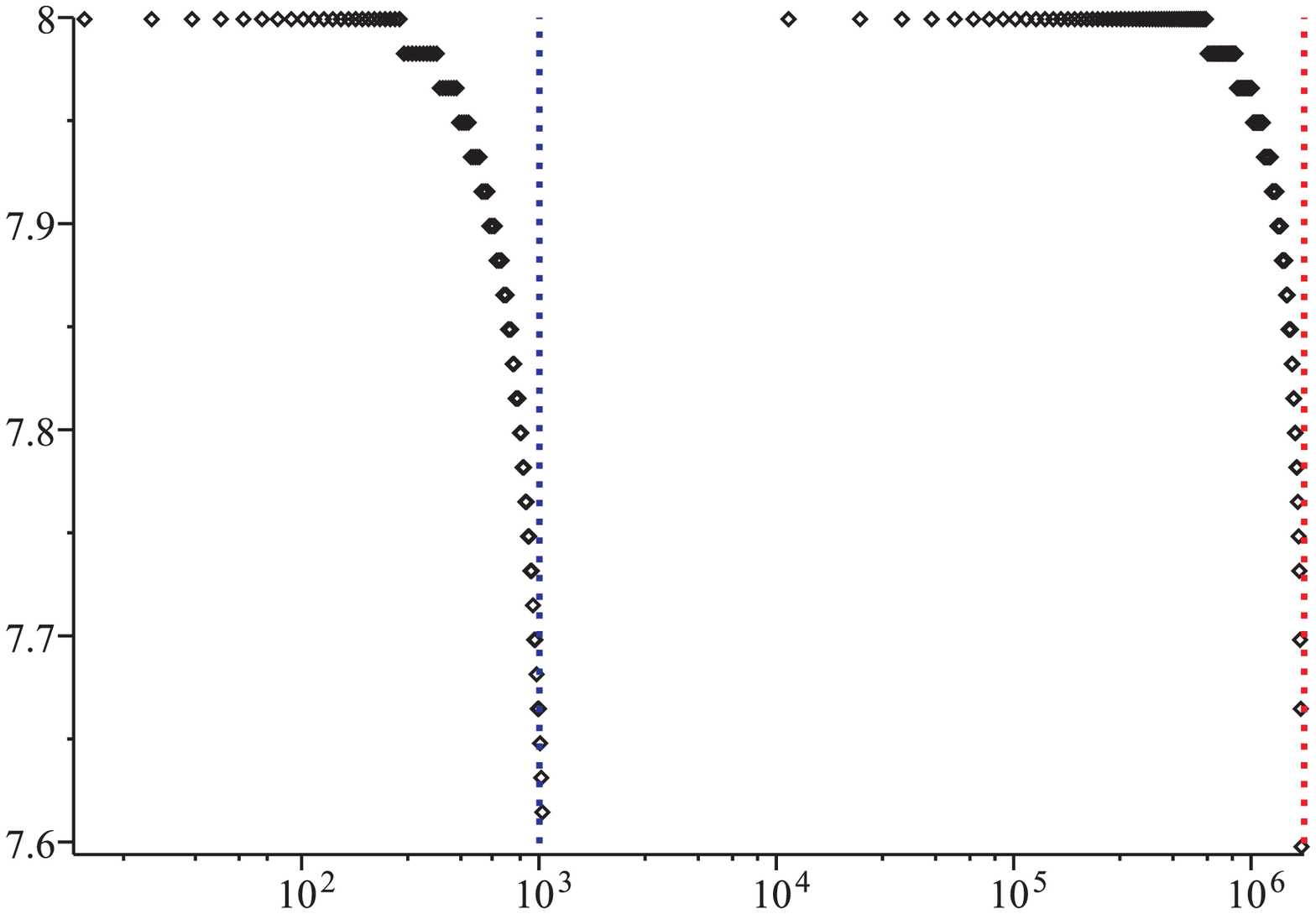}

  \caption{Upper panel: Numerical simulations of full time dependent field equation of the {kink/anti-kink for the new model} (solid line) {and for the} $\phi^4$ model (dashed line). The higher intensity of force {for the new model} is related with a stronger and consequently faster kink/anti-kink interaction. Lower panel: {Corresponding positions of the anti-kink center for the new model  (left) and for the $\phi^4$ model (right) with time.} \label{fig:new_model_snap} }
  \end{figure}

It is possible to calculate the force between a kink and an anti-kink solution, {spaced by a distance $L$ much larger than} their thickness, using the energy momentum density  $T^{01}$ component \cite{Vachaspati:2006zz}. The resulting force is given by  $|F| \propto (L/\delta)^{-6}$ for the new model solution, with $E$ given by Eq. \eqref{E_0_new}, clearly more intensive than the usual Yukawa-like $e^{-2L/\delta_s}$, obtained for $\phi^4$ model. The comparison of the two type of kink interactions was confirmed by numeric simulating the full time dependent field equation, given by Eq. \eqref{eqmotion222}  for the \eqref{phi4model} and \eqref{newmodel} potentials, with $\lambda=\tilde \lambda=a=b=1$. The simulation starts with a well separated kink/anti-kink pair at rest, for both the new-model and $\phi^4$ model solutions. The corresponding initial configuration is $\phi_0=\phi(x+x_0)-\bar\phi(x-x_0)-1$. The results are presented in the upper panel of Fig \ref{fig:new_model_snap}, where snap-shots of the two model solutions are shown for the same time-steps. It is possible to see that the time-scale associated with the two forces is very different. At the simulation time-scale the new model pair of solutions starts collapsing while the corresponding $\phi^4$ pair does not move significantly. The different time-scale of the two forces is even more clear in the lower panel of Fig. \ref{fig:new_model_snap}, where the positions of the {two anti-kink solutions for the new model (left) and for the $\phi^4$ model} (right), are plotted with time for the two models.

\section{Stability\label{stability}}

In order to investigate  behavior of the solution under small perturbations, the following general fluctuations of the scalar field are introduced: $\phi(x,t)=\phi(x)+\sum_i \eta_i(x)\cos(\omega_i \,t)$, where $\eta_i(x)$ represents a set of perturbations of  the static solution. Introducing these fluctuations, expanded to the first order in $\eta_i$, in the equation of motion, we obtain an equation of motion formally analogous to the stationary Schr\"odinger equation, dubbed Schr\"odinger-like equation:
\be\label{equation11}
\left(-\frac{d^2}{dx^2}+ U(x)\right)\eta_i=\omega_i^2 \eta_i \,\,\,\Rightarrow \,\,\, H\eta_i=\omega_i^2 \eta_i
\ee
with
\be
\label{quantipot}
U(x)=V^{(2)}_{\phi}|_{\phi=\phi(x)}.
\ee
 An eigenfunction corresponding to $\omega^2>0$ would be a localized shape change of the kink, characterizing the excitation of an internal mode \cite{dp}. The formal analogy with quantum mechanics will show that the kink acts as a potential well for the linear waves \cite{dp}, and allows us to further in this paper apply the probabilistic interpretation for the study of frequency modes and resonances. This interpretation is standard in problems of field localization in branes (see, for instance, Refs. \cite{gomes,liu}).

For models that support a superpotential function, equation \eqref{equation11} can be written as $S^\dag S \eta_i = \omega_i^2 \eta_i$, where
\be\label{SdagS}
S^\dag=-\frac{d}{dx}+W^{(2)}_{\phi} \,\,\,\,\, {\rm and} \,\,\,\,\,S=\frac{d}{dx}+W^{(2)}_{\phi}.
\ee
The non-negative value of $H=S^\dag S$ implies that the system does not support any state with a non-negative eigenvalue, ensuring the linear stability of the model. In terms of the superpotential function, Eq. \eqref{quantipot} may be written as
\be\label{U_W}
U(x)=(W^{(2)}_{\phi})^2-W^{(1)}_\phi W^{(3)}_{\phi}.
\ee
Using Eqs. \eqref{Wphi4} and \eqref{U_W}, {the Schr\"odinger-like potential} for the $\phi^4$ model can be obtained,
\be
U(x)=2b^2\tilde\lambda\left(2-3\sech^2(b\sqrt{\tilde\lambda}x)\right).
\ee
This potential is a modified reflectionless P\"oschl-Teller \cite{RosenMorse}, that allows, in addition to the zero mode, a {frequency} bound mode with eigenvalue $2\beta^2\tilde \lambda$. All the others states of the model, with $w_i^2\geq4b^2\tilde \lambda$, are unbound.

Equivalently, it is possible to obtain the {Schr\"odinger-like potential for the new model}, using the Eqs. \eqref{Wphi3} and \eqref{U_W},
\be \label{Volcano}
U(x)=12\lambda a^4 \frac{(a^2 \sqrt{\lambda} x)^2-1/4}{[1+(a^2 \sqrt{\lambda} \,x)^2]^2}.
\ee

\begin{figure}[h!]
  \includegraphics[width=7.3cm]{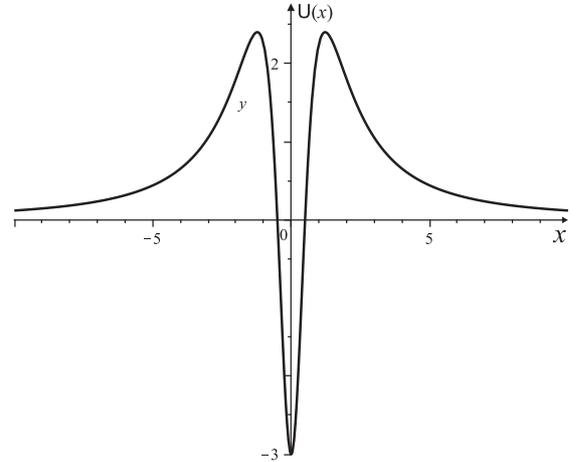}
  \caption{Profile of the {Volcano potential $U(x)$ given by} Eq. \eqref {Volcano} for $a=\lambda=1$. \label{fig:pot_quant}
 }
  \end{figure}

This potential, shown in Fig. \ref{fig:pot_quant}, is called the Volcano potential. Note it goes to zero  at infinity and exhibits two maxima points with the value $U_{\it max}=12\lambda a^4/5$, for $\phi=\pm \sqrt{15}a/5$.

Using the superpotential function, Eq.\eqref{Wphi3}, we obtain $W^{(2)}_{\phi}=-3\sqrt{\lambda} \,\phi\,\sqrt{a^2-\phi^2}$ which can be used to derive Eq.\eqref{SdagS} for the new model,
\be
S=\frac{d}{dx}-\frac{3\lambda a^4 x}{1+(a^2 \sqrt{\lambda} \,x)^2},
\ee
ensuring the non-existence of unstable bound modes with negative eigenvalues. Since the potential vanishes for large values of $x$, the only {bound mode solution} is then the zero mode:
\be
\eta_0(x)=\frac{a^3\sqrt{\lambda}}{\left(\sqrt{1+(a^2 \sqrt{\lambda} \,x)^2}\right)^3}.
\ee
All the vibrational modes with $k^2>0$ are asymptotically plane waves. The absence of positive bound modes for the new kink solution could in principle affect the rate of loss of energy by radiation in a kink-antikink collision process in comparison to the usual $\phi^4$ kink \cite{CampSch,pc}. However, note that for eigenvalues with $w^2\leq U_{\rm max}$ a possible resonance structure could localize the energy density inside the defect in a sufficiently long time to act as an effective bound state for the interacting properties of the kink. The influence of this effect for the formation of two-bounce windows \cite{AOM} in a collision processes is currently being under consideration.

 The volcano-like potentials were also investigated in the context of the braneworld scenarios where gravity is coupled to one real scalar field. For some models, the  behavior of the massive gravitational modes inside the defect are described in \cite{Gremm:1999pj,csaki,bgl,cga}.

\section{Generalization\label{generalization}}

\begin{figure}[h]
 \subfigure[Potentials]{\includegraphics[width=0.48\linewidth]{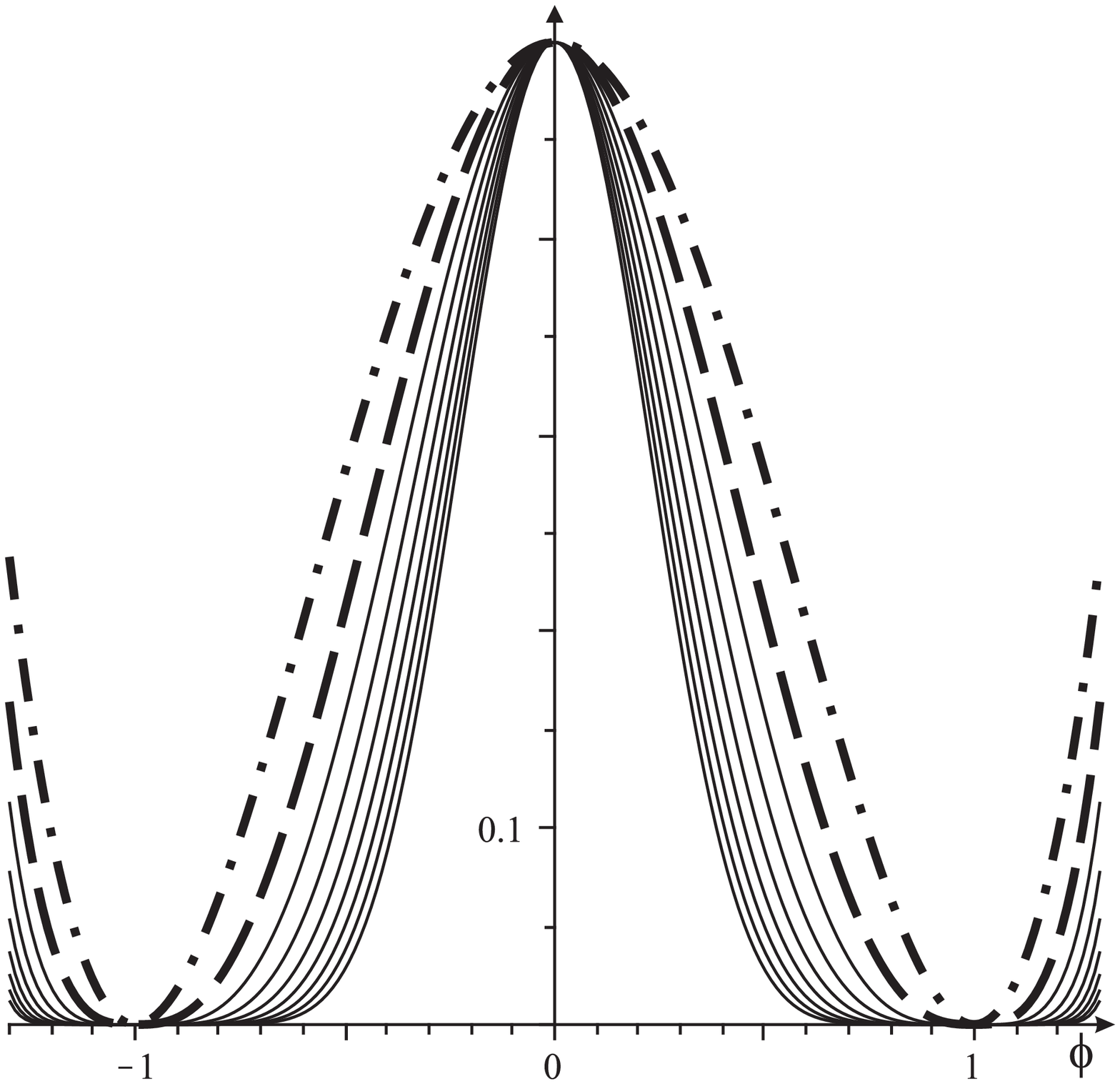}\label{seA}}
\subfigure[Static solutions]{ \includegraphics[width=0.48\linewidth]{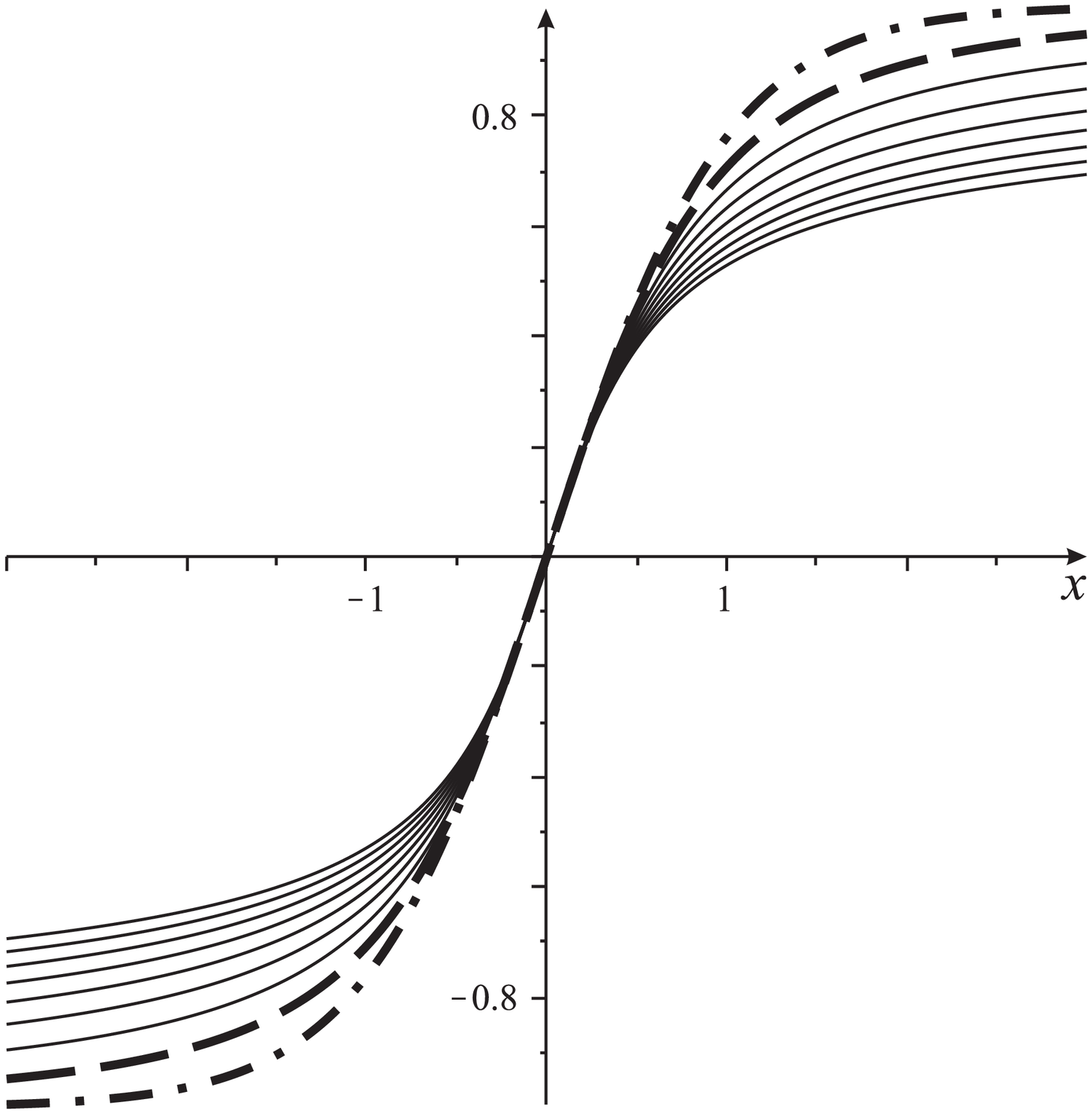}\label{seB}}  \\
 \subfigure[Energy densities]{\includegraphics[width=0.49\linewidth]{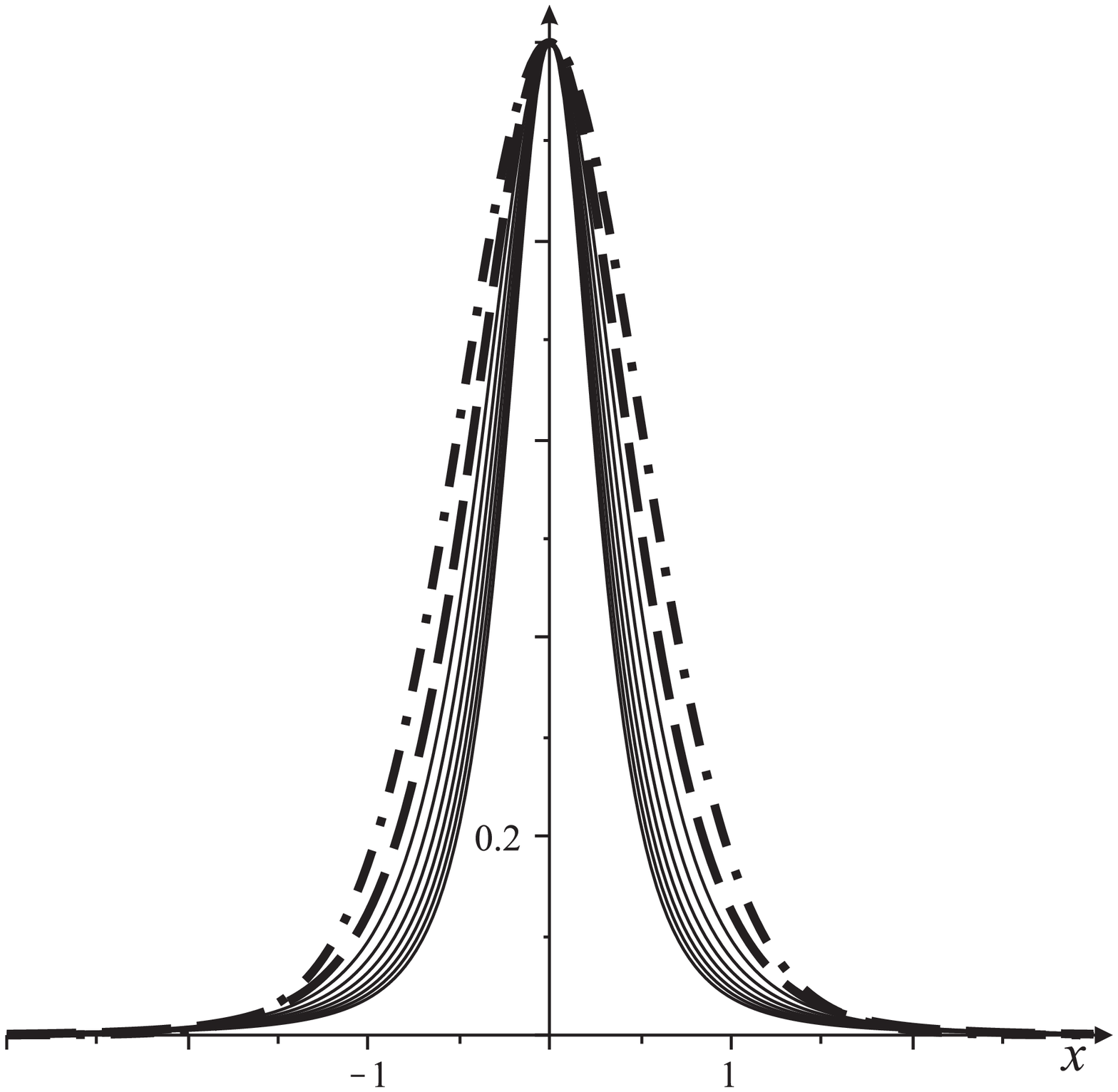}\label{seC}}
 \subfigure[Schr\"odinger-like potentials]{\includegraphics[width=0.49\linewidth]{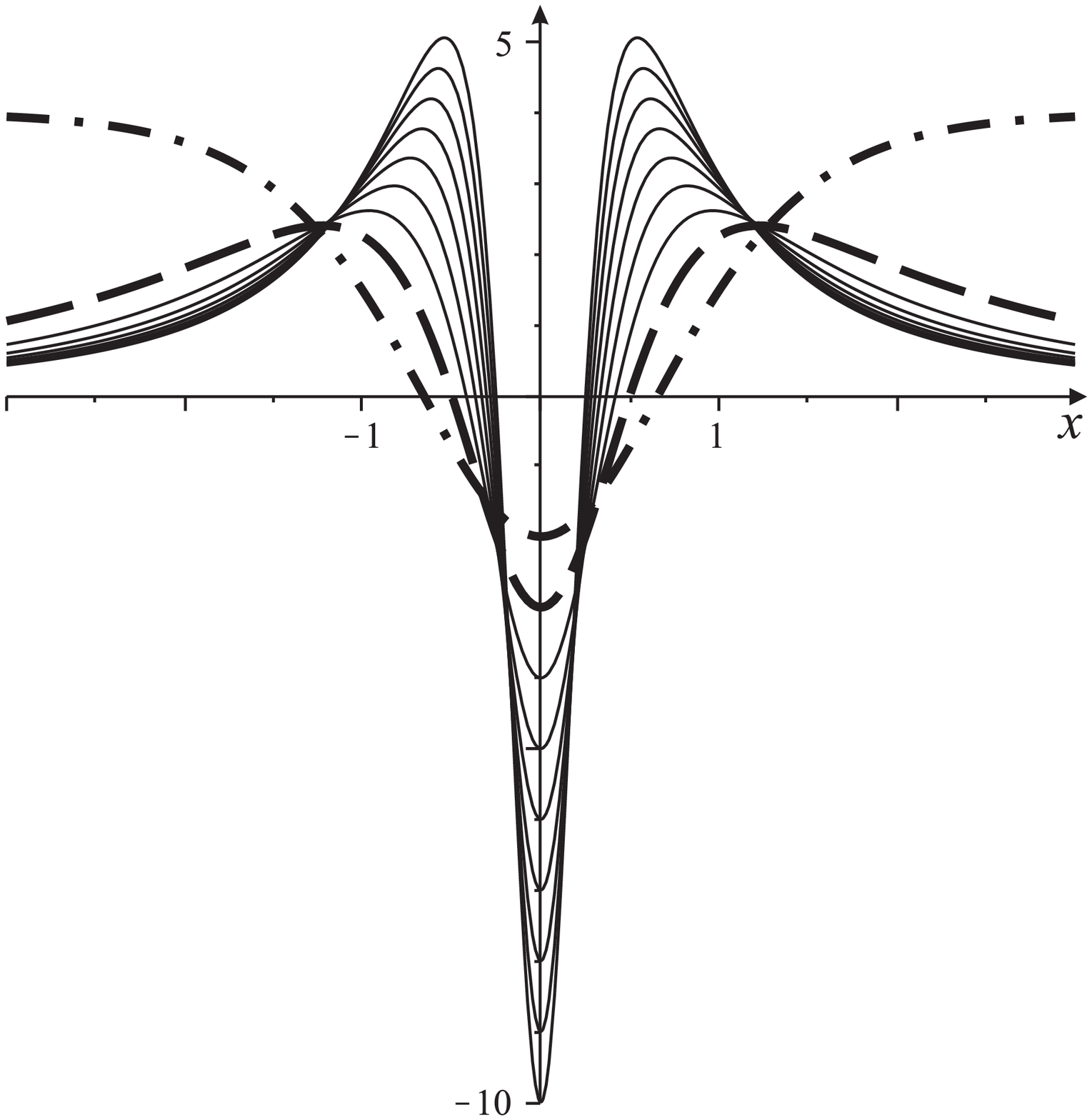}\label{seD}}
  \caption{Higly interactive kink generalized class of models, Eq.\eqref{Vn}.  The dashed-dotted and dashed lines refer to $n=1$ and $n=2$, respectively. The solid lines refer to $n=3$ to $n=9$.  We have chosen $\lambda=a=1$.\label{fig:new_model2} }
  \end{figure}

A generalization of the previous model, for solutions with a progressively slower power-law asymptotic behavior, is given by
\be\label{Vn}
V=\frac\lambda 2 |a^2-\phi^2|^{n+1},
\ee
where $n$ is an integer parameter. {A similar extension for the model of Ref. \cite{Bazeia:2002xg}  was considered in Ref. \cite{bglm} in the context of the deformation theory.}  We see that the generalized potential preserves the maximum and minima values for $\phi_{\max}=0$ and $\phi_{\min}=\pm a$. The first non-vanishing derivative orders of the potential at the minima are given by  $|V^{(n+1)}_\phi(\pm a)|=2^{n}\,(n+1)!\lambda a^{n+1}$. Note that the cases $n=1$ and $n=2$ correspond to models \eqref{phi4model} and \eqref{newmodel} respectively. The potential is plotted in Fig \ref{seA} for different values of $n$. It is possible to observe the enlargement of the minima flatness with the increasing value of n.

The corresponding generalized superpotential function $W(\phi)$, for $|\phi|<a$ is given by
\be
W(\phi)={\lambda}a^{n+1}\phi \,{F_1\left(\frac12,-\frac12\,(n+1);\frac32;\,\frac{{\phi}^{2}}{a^2}\right)}, \,\, {\rm for} \,\,|\phi|<a.
\ee
Using this potential we may calculate the total energy of the generalized solution as,
\be
E_n=a^{n}{\sqrt{\pi\,\lambda}} \frac{\Gamma\left(\frac32+\frac{n}{2}\right)}{\Gamma\left(2+\frac{n}{2}\right)}.
\ee
Considering the generalized equation of motion,
\be\label{general_eq}
\frac{d^2\phi}{dx^2}=- (n+1)\lambda\phi(a^2-\phi^2)|a^2-\phi^2|^{n-1},
\ee
we determine the numeric solutions of the generalized class of models, and their corresponding energy densities, plotted in Fig \ref{seB} and \ref{seC}, for different values of $n$. Is is clear that the solutions and energy densities asymptotic behavior slow down with an increasing $n$.

 Using a perturbative method we also calculated the analytic asymptotic behavior of the generalized model solutions and corresponding energy densities, for different values of $n>1$,

\bes
 \ben
 \phi_n(x)\!&=&\! a \left[1 - \frac{({n-1})^{-\frac{{2}}{n-1}}}2\left({\frac{\delta_n}{2x}}\right)^{\frac{2}{n-1}}+\ldots \right]\\
 \rho_n(x)\!&=&\! \!a^{2n+2}\! \left(n-1\right)^{-\frac{2(n+1)}{n-1}} \!\!\left({\frac{\delta_n}{2x}}\right)^{\frac{2(n+1)}{n-1}}
 \!\!+\!\ldots
 \een
\ees
where we define $\delta_n=2(\sqrt{\lambda}a^{n})^{-1}$ as the $n^{th}$ solution thickness. Note that the asymptotic behavior of these solutions is included in power-law behavior presented in Eq. \eqref{powerlaw} at section \ref{Intro}.

\begin{figure}[ht]
  \includegraphics[width=7cm]{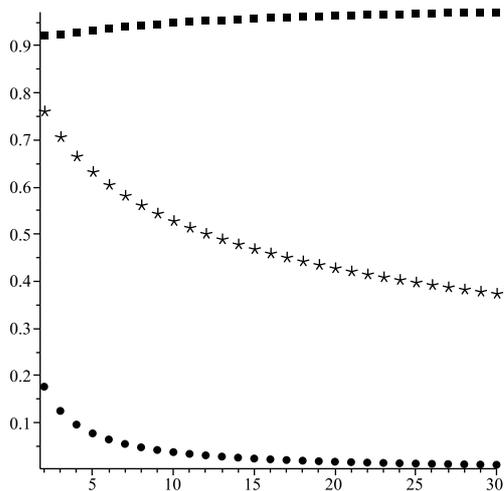}
  \caption{\label{fig:pot_thickness} The ratio $\phi_n(\delta_n/2)/\phi_n(\infty)$ (asterisks), the energy density of the solutions $\rho_n(\delta_n/2)$ (circles) for $x=\delta_n/2$ and the energy ratio given by Eq. \eqref{ratio} (boxes), for different values of $n$}.
  \end{figure}

{For a better} illustration of the progressively slower asymptotic  behavior of the solutions we have also plotted the ratio between the value of $\phi$ for $x=\delta_n/2$ and its value at the minima, $\phi_n(\delta_n/2)/\phi_n(\infty)$, in Fig. \ref{fig:pot_thickness} (asterisks). A loss of accuracy in the thickness definition, $\delta_n$, could be associated with an increasing of the solutions spread, however this behavior is accompanied by an increasing localization of their energy density, as shown in Fig.\ref{seB}. This result is also confirmed in Fig.\ref{fig:pot_thickness} where the solutions energy densities (circles) are plotted for $x=\delta_n/2$, and moreover in Fig \ref{seC} (boxes) where the ratio
\be\label{ratio}
\frac{\displaystyle \int_{x=-{\delta_n}/{2}}^{x={\delta_n}/{2}} \rho_n(x) dx}{\displaystyle \int_{x=-{\infty}}^{x={\infty}} \rho_n(x) dx},
\ee
is shown, corroborating the increasing of the energy density localization with $n$.

\begin{figure}[ht]
  \includegraphics[width=7cm]{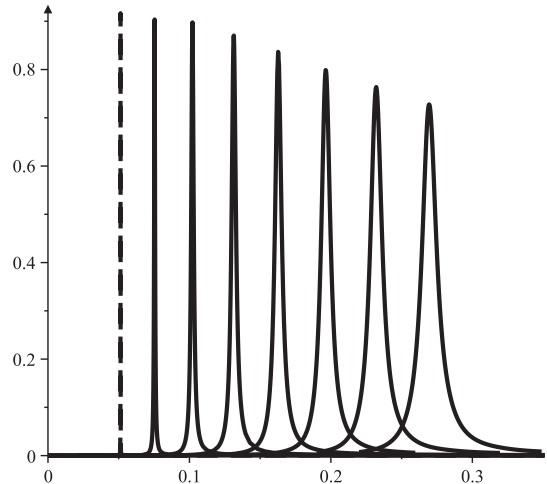}
  \caption{Single resonance peaks characterizing the relative  probability $P_{rel}(0)$ for finding {frequency} modes near to the center of the kink, as a function of the  frequency $\omega$ of the mode. The peaks are for $n=2$ to $n=9$. \label{fig:ress} The lower is $n$, the thinner is the corresponding resonance peak. }
  \end{figure}

Finally in Fig. \ref{seD} the Schr\"odinger-like potentials $U_n(x)$ are plotted for different values of $n$. The Volcano potential well tapers with the increase of $n$ and the maxima points in $\phi=\pm \sqrt{3}a/\sqrt{2n-1}$ with $U_{max}=2^{n-1}n (n-2)^{n-2}(2n-1)^{2-n}\lambda a^n$.  {The growth of $U_{max}$ with $n$  may be associated with occurrence of resonances with higher frequencies when $n$ is increased. In order to confirm this hypothesis, we considered even parity states and applied the well-known Numerov method \cite{num} (for  application of the method to brane problems see ref. \cite{bgl} and references therein) for finding the {frequency} modes $\eta_m(x)$. We want to compare the relative probability $P_{rel}(0)$ for finding a {frequency} mode around the kink center located at $x=0$. This justifies choosing a fixed control distance $x_c$ and defining, for a  known numerical Schr\"odinger-like potential for $-x_{max}\le x\le x_{max}$
\begin{equation}
P_{rel}(0)=\frac{\displaystyle \int_{x=-x_c}^{x=x_c} \eta_m(x) dx}{\displaystyle \int_{x=-{x_{max}}}^{x={x_{max}}} \eta_m(x) dx},
\end{equation}
{where $x_{max}$ allows for a sufficiently large number of oscillations in the range $x_{max}<x<x_{max}$ for the wavefunctions $\eta_m(x)$.}
Fig. \ref{fig:ress} shows the resonance peaks $P_{rel}(0)$ found for $n=2$ to $n=9$, as a function of the frequency $\omega$ of the mode.  The first thin peak is for $n=2$. The sequence of peaks in the figure follow the increasing sequence of $n$. Note that each peak characterizes a resonance centred at a particular value of $m$. The whole sequence shows that the increasing of $n$ turns the resonance peak broader. The first thinner and higher peak for $n=3$ is the first modification to the $\phi^4$ model that has a bound {frequency} state. The modification for $n=3$ trades the bound state for $n=2$ by a quasi-normal state with lifetime inversely proportional to the peak thickness. The growth of $n$ leads to higher {frequency} resonant states, agreeing with our discussion after Eq. (\ref{ratio}).}

\section{Ending Comments\label{end}}

In this work we have introduced a new real scalar field model that admits kink solutions exhibiting a power-law asymptotic behavior, instead of the usual exponential one. This is a consequence of having a potential with a null second order derivative at the minima.  We have shown that this particular asymptotic behavior of the solution leads to a much stronger kink/anti-kink interaction, when the same thickness and energy-scale of the defects is considered, than in the standard models.

We have further generalized the model to a class of models admitting $n$ vanishing derivative orders of the potential at the minima. It was found that, despite the {increasing} of the solutions spread with $n$, their corresponding energy density becomes more localized when $n$ is incremented.

 {Stability analysis led to a Schr\"odinger-like equation where bound modes are absent. This} may be related with interesting properties of defect collisions. The two-bounce effect \cite{CampSch,AOM}  appearing in the collision of $\phi^4$ model defects requires the transference of a portion of kinetic collision energy to the internal positive bound mode of the solutions.
Moreover we have shown that the Schr\"odinger-like potential of our model is the Volcano potential. The inexistence of {frequency} bound modes in this potential suggests that no two-bounce effect is observed in this case, however a resonance effect may occur inside the defects during the collision, leading to some peculiar behavior. This is an interesting future work.

 {Moreover the embedding of these kink solutions in four spacetime dimensions generates a domain wall network with a particular dynamics, which could have interesting applications in Cosmology or in Condensed Matter in the non-relativistic limit. Note that different energy factors have to be considered in dynamics of domain wall networks, not only the kinetic energy or the radiation energy losses but also the wall curvature needs to be put into play. Possible implications of this class of models in the dynamics of domain wall networks is also an interesting future study.}

Another possible relevant application for this class of models is in the braneworld scenario. Indeed, as one knows, using the first-order formalism the extension for $(4,1)$ dimensions of this kink-like solutions is straightforward \cite{DeWolfe}. Gravity and particle localization in branes have been studied extensively in the literature, with the main focus of looking for resonance phenomena \cite{gomes,liu, liu-ads,cga,cga2,liu-nonmin}. For the case presented here, the special characteristics of perturbations around the kink solution suggests that similar studies is also a very interesting issue to be further considered.

\section*{Acknowledgements}
We would like to thank CAPES, CNPq, FAPEMA, Brasil, and FCT project CERN/FP/116358/2010, Portugal, for partial financial support and D. Bazeia, L. Losano and J. G. Ramos for useful discussions. A. R. Gomes thanks ITP-UH and  R. Menezes thanks the Centro de F\'\i sica do Porto for all hospitality during this work.


\begin{thebibliography}{99}

\bibitem{Vachaspati:2006zz}
  T.~Vachaspati, {\it Kinks and domain walls: An introduction to classical and quantum solitons}, Cambridge University Press, Cambridge, England, 2006.

\bibitem{VSh}A. Vilenkin and E.P.S. Shellard, {\it Cosmic Strings and Other Topological Defects}, Cambridge University Press, Cambridge, England, 1994.

\bibitem{ManSut}
N. Manton and P. Sutcliffe, {\it Topological Solitons}, Cambridge University Press, Cambridge, England, 2004.

\bibitem{GW} J.A. Glazier and D. Weaire, Journal of Physics Condensed Matter {\bf 4}, 1867 (1992);

\bibitem{Provatas} N. Provatas and K. Elder, {\it Phase-Field Methods in Materials Science and Engineering}, Wiley-VCH, 2010;

\bibitem{WH} D. Weaire and R. Hutzler, {\it The Physics of Foams}, Oxford University Press, Oxford, 2000;


\bibitem{Travasso} R.D.M. Travasso, M. Castro, and J.C.R.E. Oliveira, Philosophical Magazine {\bf 91}, 183 (2011);


\bibitem{loginova}  I.S. Loginova and H.M. Singer, Rep. Prog. Phys. {\bf 71}, 106501 (2008);

\bibitem{bray} A.J. Bray, Adv. Phys., {\bf 51}, 481 (2002)


\bibitem{wall1} M. Bucher and D. N. Spergel, Phys. Rev. D {\bf 60}, 043505

\bibitem{wall2}  P.P. Avelino, C.J.A.P. Martins, J. Menezes, R. Menezes, and J. C. R. E. Oliveira, Phys. Rev. D {\bf73}, 123519 (2006).

\bibitem{wall3}  P.P. Avelino, C.J.A.P. Martins, J. Menezes, R. Menezes, and J. C. R. E. Oliveira, Phys. Rev. D{\bf 78}, 103508 (2008).

\bibitem{wall4} P.P. Avelino, C.J.A.P. Martins, and J.C.R.E. Oliveira, Phys. Rev. D{ \bf 72}, 083506 (2005).

\bibitem{Avelino:2010qf} P.~P.~Avelino, R.~Menezes and J.~C.~R.~E.~Oliveira, Phys.\ Rev.\ E {\bf 83}, 011602 (2011).

\bibitem{acampos}A. Campos, Phys.Rev.Lett. 88 (2002) 141602.

\bibitem{DeWolfe} O.~DeWolfe, D.~Z.~Freedman, S.~S.~Gubser and A.~Karch, Phys.\ Rev.\ D {\bf 62}, 046008 (2000).

\bibitem{Toharia:2007xe} M.~Toharia and M.~Trodden, Phys.\ Rev.\ Lett.\  {\bf 100}, 041602 (2008),

\bibitem{Basar:2008im}G.~Basar and G.~V.~Dunne, Phys.\ Rev.\ Lett.\  {\bf 100}, 200404 (2008).

\bibitem{Melfo:2011ev}A.~Melfo, R.~Naranjo, N.~Pantoja, A.~Skirzewski and J.~C.~Vasquez, Phys.\ Rev.\ D {\bf 84}, 025015 (2011).

\bibitem{VBA} A. Vanhaverbeke, A. Bischof, and R. Allenspach, Phys.\ Rev.\ Lett. {\bf 101}, 107202 (2008)

\bibitem{ABC} A.T. Avelar, D. Bazeia, and W.B. Cardoso, Phys. Rev. E {\bf 79}, 025602(R) (2009).

\bibitem{Bazeia:2006pj}D.~Bazeia, M.~A.~Gonzalez Leon, L.~Losano and J.~Mateos Guilarte, Phys.\ Rev.\ D {\bf 73}, 105008 (2006)

\bibitem{AlonsoIzquierdo:2010ze} A.~Alonso-Izquierdo, M.~A.~G.~Leon, J.~M.~Guilarte and M.~de la Torre Mayado, JHEP {\bf 1008}, 111 (2010).

\bibitem{AKL} M.J. Ablowitz, M.D. Kruskal, and J.F. Ladik, SIAM J. Appl. Math., {\bf 36}, 428 (1979).

\bibitem{CampSch} D. K. Campbell, J. S. Schonfeld, and C. A. Wingate, Phys. D, {\bf 9}, 1 (1983).

\bibitem{pc} M. Peyrard, D.K. Campbell, Physica 9D (1983) 33.

\bibitem{AOM} P. Anninos, S. Oliveira, and R.A. Matzner, Phys. Rev. D, {\bf 44}, 1147 (1991).

\bibitem{doi} Y. Doi, Phys. Rev. E {\bf 68}, 066608 (2003).

\bibitem{Dorey:2011yw} P.~Dorey, K.~Mersh, T.~Romanczukiewicz and Y.~Shnir, Phys.\ Rev.\ Lett.\  {\bf 107}, 091602 (2011).

\bibitem{Babichev:2006cy} E.~Babichev, Phys.\ Rev.\ D {\bf 74}, 085004 (2006).

\bibitem{Bazeia:2007df} D.~Bazeia, L.~Losano, R.~Menezes and J.C.R.E.~Oliveira, Eur.\ Phys.\ J.\ C {\bf 51}, 953 (2007).

\bibitem{Adam:2007ij} C.~Adam, J.~Sanchez-Guillen and A.~Wereszczynski, J.\ Phys.\ A A {\bf 40}, 13625 (2007) [Erratum-ibid.\ A {\bf 42}, 089801 (2009)].

\bibitem{Bazeia:2008tj} D.~Bazeia, L.~Losano and R.~Menezes, Phys.\ Lett.\ B {\bf 668}, 246 (2008).

\bibitem{Bazeia:2011sb} D.~Bazeia, J.D.~Dantas, A.R.~Gomes, L.~Losano and R.~Menezes, Phys.\ Rev.\ D {\bf 84}, 045010 (2011).

\bibitem{Adam:2011jt} C.~Adam and J.~M.~Queiruga, Phys.\ Rev.\ D {\bf 84}, 105028 (2011).

\bibitem{Bog}  E. B. Bogomol'nyi, Sov. J. Nucl. Phys. {\bf 24}, 449 (1976).

\bibitem{PS} M. K. Prasad and C. M. Sommerfield, Phys. Rev. Lett. {\bf 35}, (1975) 760.

\bibitem{Vilenkin:1994pv} A.~Vilenkin, Phys.\ Rev.\ Lett.\  {\bf 72}, 3137 (1994).

\bibitem{Dinda} P. Tchofo Dinda, Phys.\ Rev.\ B {\bf 46}, 12012 (1992).

\bibitem{Bazeia:2002xg} D.~Bazeia, L.~Losano and J.M.C.~Malbouisson, Phys.\ Rev.\ D {\bf 66}, 101701 (2002).

\bibitem{dp} T. Dauxois, M. Peyrard, Cambridge, UK: Univ. Pr. (2006) 422p.

\bibitem{gomes} C.A.S. Almeida, R. Casana, M.M. Ferreira Jr., A.R. Gomes, Phys. Rev. D79 (2009) 125022.

\bibitem{liu} Y.X. Liu, J. Yang, Z.H. Zhao, Chun-E Fu and Y.S. Duan, Phys. Rev. D80 (2009) 065019.

\bibitem{RosenMorse} N. Rosen and P.M. Morse, Phys. Rev. {\bf 42} 210 (1932)

\bibitem{Gremm:1999pj}M.~Gremm, Phys.\ Lett.\ B {\bf 478}, 434 (2000).

\bibitem{csaki} C. Csaki, J. Erlich, T. J. Hollowood, Phys. Rev. Lett {\bf 84}, 5932 (2000).

\bibitem{bgl} D. Bazeia, A. R. Gomes, L. Losano, Int. J. Mod. Phys. A {\bf 24}, 1135 (2009).

\bibitem{cga}  W.T. Cruz, A.R. Gomes, C.A.S. Almeida, Europhys. Lett. {\bf 96}, 31001 (2011).

\bibitem{bglm} D. Bazeia, M. A. Gonz\'alez Leon, L. Losano, J. Mateos Guilarde, Europhys. Lett. {\bf 93}, 41001 (2011).

\bibitem{num} B. V. Numerov, Roy. Ast. Soc. Monthly Notices {\bf 84}, 592 (1924).

\bibitem{liu-ads} Yu-Xiao Liu, Heng Guo, Chun-E Fu, Ji-Rong Ren, JHEP 1002 (2010) 080.

\bibitem{cga2}  W.T. Cruz, A.R. Gomes, C.A.S. Almeida, Eur. Phys. J. C {\bf 71}, 1790 (2011).

\bibitem{liu-nonmin} Heng Guo, Yu-Xiao Liu, Zhen-Hua Zhao, Feng-Wei Chen, arXiv:1106.5216 [hep-th].


\end{thebibliography}
\end{document}